\documentclass[conference]{IEEEtran}
\usepackage{multirow}
\usepackage[pdftex]{graphicx}
\usepackage[ruled]{algorithm2e}
\usepackage{algorithmic}
\usepackage[small]{caption}
\usepackage{float}
\usepackage{amssymb,amsmath}
\usepackage{graphicx}
\usepackage{subfigure}

\usepackage{xspace}
\usepackage{fancyhdr}
\usepackage{amssymb,amsmath}
\usepackage{textcomp} 
\usepackage{epstopdf}
\usepackage{bbding}
\usepackage{cite} 
\usepackage{color}
\usepackage{pifont}

\begin{document}
%
\title{\LARGE Integration of Backscatter Communication with Multi-cell NOMA:\\ A Spectral Efficiency Optimization under Imperfect SIC}
\author{Wali Ullah Khan$^1$, Eva Lagunas$^1$, Asad Mahmood$^1$, Zain Ali$^2$, Symeon Chatzinotas$^1$,\\  Bj\"orn Ottersten$^1$, Octavia A. Dobre$^3$, \\$^1$Interdisciplinary Centre for Security, Reliability and Trust (SnT), University of Luxembourg\\
$^2$Department of Electrical and Computer Engineering, University of California, Santa Cruz, USA\\
$^3$Memorial University, St. John's, NL A1B 3X9, Canada \\
\{waliullah.khan, asad.mahmood, eva.lagunas, symeon.chatzinotas, bjorn.ottersten\}@uni.lu\\
zainalihanan1@gmail.com; odobre@mun.ca\thanks{This work was supported by Luxembourg National Research Fund (FNR) under the CORE project ------.}

}%

\maketitle
\begin{abstract}
Future wireless networks are expected to connect large-scale low-powered communication devices using the available spectrum resources. Backscatter communications (BC) is an emerging technology towards battery-free transmission in future wireless networks by leveraging ambient radio frequency (RF) waves that enable communications among wireless devices. Non-orthogonal multiple access (NOMA) has recently drawn significant attention due to its high spectral efficiency. The combination of these two technologies can play an important role in the development of future networks. This paper proposes a new optimization approach to enhance the spectral efficiency of nonorthogonal multiple access (NOMA)-BC network. Our framework simultaneously optimizes the power allocation of base station and reflection coefficient (RC) of the backscatter device in each cell under the assumption of imperfect signal decoding. The problem of spectral efficiency maximization is coupled on power and RC which is challenging to solve. To make this problem tractable, we first decouple it into two subproblems and then apply the decomposition method and Karush-Kuhn-Tucker conditions to obtain the efficient solution. Numerical results show the performance of the proposed NOMA-BC network over the pure NOMA network without BC.
\end{abstract}



\section{Introduction}
\IEEEPARstart{B}{eyond} 5G technologies are expected to improve connection density, latency, energy consumption, and data reliability \cite{9040264}. Power domain nonorthogonal multiple access (NOMA) can support a huge number of wireless devices over the limited spectrum resources with high data rates and low transmission delay in B5G networks \cite{9479745}. Recently, a new wireless primitive called backscatter communications (BC) has been introduced that enables interactive devices which compute and communicate without any batteries. According to its working principle, BC enables wireless communications by leveraging the ambient signals all around us, instead of generating our own \cite{9261963}. In particular, the backscatter device (also called tag) (BD) can send a message to the users (also called readers) by either absorbing or reflecting signals from the TV tower, WiFi access point, and cellular base station (BS) \cite{9363336}. Thus, combining BC with NOMA in a communication system can significantly enhance energy and spectral efficiency. 

The performance of BC systems in NOMA networks have recently been studied in the literature. For example, the authors in \cite{chen2021backscatter,nazar2021ber,li2020secrecy,zhang2019backscatter} have investigated the outage probability, throughput, ergodic capacity, bit error rate, and intercept probability of the NOMA-BC networks. Moreover, researchers have also proposed various resource allocation schemes in NOMA-BC networks. For instance, Khan {\it et al.} \cite{93285055} have optimized the power of BS and reflection coefficient (RC) of BD to enhance the spectral efficiency (SE) of NOMA-BC under imperfect signal decoding. In another study, Yang {\it et al.} \cite{8851217} have optimized the time and RC of BD for the system max-min throughput of NOMA-BC. The research of Chen {\it et al.} \cite{9122620} has optimized the transmit power of the BS and RC of BD to improve the system ergodic capacity. Moreover, the study of \cite{9223730} has optimized the transmit power of the BS and RC of BD to maximize the energy efficiency of the system. Of late, Khan {\it et al.} \cite{9345447} have computed the power of BS and roadside units to maximize the SE of vehicular network. 

The existing works \cite{chen2021backscatter,nazar2021ber,li2020secrecy,zhang2019backscatter,93285055,8851217,9122620,9223730} have considered two-user and single-cell scenarios that are very simple and impractical. Besides, most of the works \cite{8851217,9122620,9223730} assume perfect signal decoding which is impractical. Further, several works consider performance analysis, they do not consider resource optimization. However, allocating network resources efficiently is very crucial in the development of large-scale wireless networks. Based on the recent literature, the problem of SE maximization for NOMA-BC that simultaneously optimizes the power allocation of BS and RC of BD under imperfect signal decoding at the receiver has not yet been investigated. To fill this gap, this paper proposes more practical optimization framework to maximize the SE of the system while ensuring the user quality of services. In particular, we consider inter-cell interference due co-channel and intra-cell interference due to NOMA users. Moreover, we assume consider that users cannot always decode their signals, hence we consider imperfect signal decoding. We formulate a maximization problem of the system SE subject to various practical constraints. Our proposed problem is non-convex which makes it very challenging to obtain the optimal solution. Thus, we first transform it and then achieve a suboptimal yet efficient solution. Specifically, the decomposition method and Karush-Kuhn-Tucker (KKT) conditions are adopted to calculate the closed-form expressions. Numerical results confirm the benefits of our NOMA-BC network over the traditional pure NOMA network. The remaining of this paper can be structured as follow. Section II provides the proposed system model of NOMA-BC and problem formulation of spectral efficiency maximization. Section III presents our proposed solution including problem transformation and closed-form expressions. Section IV provides and discusses the numerical results of the solution provided in Section III. Finally, Section V concludes this paper with some future research directions.
\begin{figure*} [!t]
\begin{align}
A&=P_{j}^2 (-F_{n,j} F_{m,j} H_{m,j} (1 + \varLambda_{n,j}) (Q_{n,j} + H_{n,j} P_{j}) + F_{n,j}H_{m,j}^2 (1 + \varLambda_{n,j}) (Q_{n,j} + H_{n,j} P_{j})\nonumber\\&+  F_{n,j} F_{m,j} H_{n,j}(1 +\varLambda_{m,j}) (Q_{m,j} + H_{m,j} P_{j}) -  F_{m,j} H_{n,j}^2 (1 + \varLambda_{m,j}) (Q_{m,j} + H_{m,j}P_{j})),\tag{13}\label{13}\\
B&=P_{j} (Q_{n,j} + H_{n,j} P_{j}) (-F_{n,j} Q_{m,j} (-2 H_{m,j} (1 + \varLambda_{n,j}) + F_{m,j}(2 + \varLambda_{m,j}))\nonumber\\& + F_{n,j} F_{m,j} H_{m,j} (\varLambda_{n,j} - \varLambda_{m,j}) P_{j} + 2F_{m,j} H_{n,j} (1 + \varLambda_{m,j}) (Q_{m,j} + H_{m,j} P_{j}))\tag{14}\label{14}\\
S&=(Q_{n,j} + H_{n,j} P_{j}) (F_{n,j} Q_{m,j}^2 (1 + \varLambda_{n,j}) +F_{m,j} (-Q_{n,j} (1 +\varLambda_{m,j}) (Q_{m,j} + H_{m,j} P_{j})\nonumber\\& + P_{j} (F_{n,j} Q_{m,j} (1 + \varLambda_{n,j}) - H_{n,j}(1 + \varLambda_{n,j}) (Q_{m,j} + H_{m,j} P_{j}))))\tag{15}\label{15}
\end{align}\hrulefill
\end{figure*}
\section{Proposed System and Formulated Problem}
\begin{figure}[t]
\centering
\includegraphics [width=0.45\textwidth]{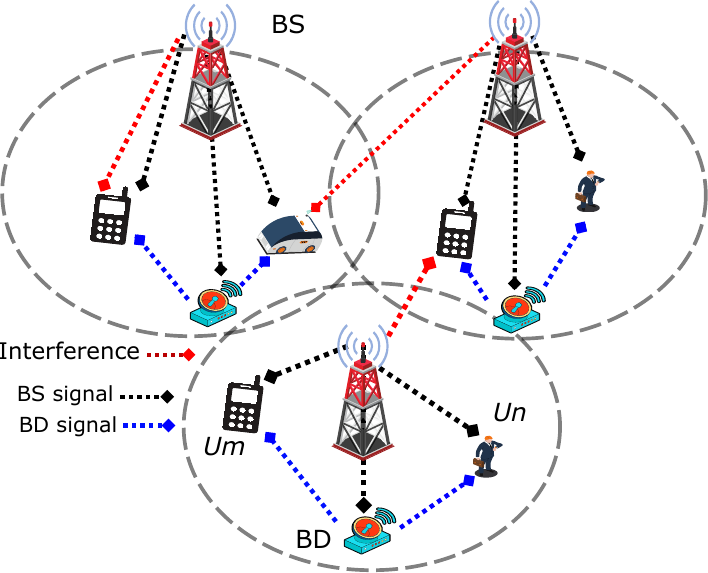}
\caption{System model}
\label{blocky}
\end{figure}
We consider a NOMA-BC network which consists of multi-cell, where each cell has single BS. To enhance the SE, we assume that all BSs resuse the same spectrum resources. Let $\mathcal C=\{C_j|j=1,2,3,\dots,J\}$ denotes the set of cells. We consider that $C_j$ serves two cellular users ($U_n$ and $U_m$) at a given time using NOMA protocol \cite{khairy2021data}. If the transmission power of $C_j$ is $P_j$ and its allocation coefficient for $U_n$ and $U_m$ is denoted as $\varphi_{n,j}$ and $\varphi_{m,j}$, then the signal that $C_j$ sends to $U_n$ and $U_m$ can be defined as $x_j=\sqrt{P_j\varphi_{n,j}}x_{n,j}+\sqrt{P_j\varphi_{m,j}}x_{m,j}$. During the communication process, a BD in the coverage area of $C_j$ (denoted as $D_j$) also detects $x_j$, harvest energy from it, modulates a signal $e_j(t)$, and then reflect it towards $U_n$ and $U_m$ with a power of $\beta_j$, where $\mathbb E[|e_j(t)|^2]=1$. Let us model the channel from $C_j$ to $U_n$ and $U_m$ as $h_{n,j}=\bar{h}_{n,j}d^{-\delta/2}_{n,j}$ and $h_{m,j}=\bar{h}_{m,j}d^{-\delta/2}_{m,j}$, where $\bar{h}_{\iota,j}\sim\mathcal{CN}(0,1)$, $\iota\in\{n,m\}$ denotes the coefficient of Rayleigh fading, $d_{\iota,j}$ is the distance between $C_j$ and $U_\iota$ and $\delta$ represents the pathloss exponent \cite{zamani2018energy}. This work assumes a perfect channel state information, thus, the signals that $U_n$ and $U_m$ receive from $C_j$ can be stated as
\begin{align}
y_{n,j}&=\sqrt{h_{n,j}}x_j+\sqrt{\beta_{j}g_{n,j}(h_{b,j}}x_{n,j})e_j(t)\nonumber\\&+\sum\limits_{j'=1, j'\neq j}^J\sqrt{h_{n,j'}}x_{j'}+\varpi_{n,j},\label{1}\\
y_{m,j}&=\sqrt{h_{m,j}}x_{j}+\sqrt{\beta_{j}g_{m,j}(h_{b,j}}x_{m,j})e_j(t)\nonumber\\&+\sum\limits_{j'=1, j'\neq j}^J\sqrt{h_{m,j'}}x_{j'}+\varpi_{m,j},\label{2}
\end{align}
where the first segment in (\ref{1}) and (\ref{2}) is the received signal from $C_j$, the second segment is the reflected signal from $D_j$, the third segment represents the inter-cell interference, and the fourth segment denotes the noise with variance $\sigma^2$. The term $g_{n,j}$ and $g_{m,j}$ represent the channels from $D_j$ to $U_n$ and $U_m$, respectively. Moreover, $h_{n,j'}$ and $h_{m,j'}$ are the channel gains at $U_n$ and $U_m$ from $C_{j'}$. Here we assume that assume that $U_n$ experiment better channel conditions than $U_m$ (i.e., $|h_{n,j}|^2>|h_{m,j}|^2$). Based on (\ref{1}) and (\ref{2}), the signal-to-interference-plus-noise ratio (SINR) at $U_n$ and $U_m$ can be expressed as
\begin{align}
\gamma_{n,j}=\frac{P_j\varphi_{n,j}(|h_{n,j}|^2+\varTheta_{n,j})}{P_j\varphi_{m,j}|h_{n,j}|^2\alpha+I_{n,j'}+\sigma^2}, \\
\gamma_{m,j}=\frac{P_j\varphi_{m,j}(|h_{m,j}|^2+\varTheta_{m,j})}{P_j\varphi_{n,j}(|h_{m,j}|^2+\varTheta_{m,j})+I_{m,j'}+\sigma^2},
\end{align}
where $\varTheta_{n,j} =\beta_{j}|h_{b,j}|^2|g_{n,i}|^2$ and ${\varTheta_{m,j} =\beta_{j}|h_{b,j}|^2|g_{m,i}|^2 }$ are the received signals from BD.\footnote{The inter-cell interference due to BC is almost negligible because of short-range transmission; thus, we can safely ignore it in our proposed framework.} The term $I_{\kappa,j'}=|h_{\kappa,j'}|^2\sum_{j'=1}^{J}P_{j'}$ is the inter-cell interference and $\alpha$ represents the imperfect successive interference cancellation (SIC) parameter which can be defined as $\alpha=\mathbb E[|x_{\kappa,j}-\tilde{x}_{\kappa,j}|^2]$, where $x_{\kappa,j}-\tilde{x}_{\kappa,j}$ is the difference between the original and the estimated signals. The objective of this work is to enhance the SE of the NOMA-BC while satisfy the QoS of all users. In particular, we simultaneously optimize $P_j$, $\varphi_{n,j}$, $\varphi_{m.j}$ and $\beta_{j}$ in each cell under imperfect signal decoding. Mathematically, it can be formulated as
\begin{alignat}{2}
\underset{{\boldsymbol{\varphi_j}, P_j,\beta_{j}}}{\text{max}}& \sum\limits_{j=1}^J\{\log_2(1+\gamma_{n,j})+\log_2(1+\gamma_{m,j})\}\label{5}\\
 s.t.\ Z1:\ &    P_j\varphi_{n,j}\left(|h_{n,j}|^2+\varTheta_{n,j}\right)\geq\left(2^{R_{req}}-1\right)\nonumber\\ & \left(|h_{n,j}|^2P_j\varphi_{m,j}\alpha+I_{n,j'}+\sigma^2\right), \forall j, \nonumber\\
 Z2:\ &  P_j\varphi_{m,j}\left(|h_{m,j}|^2+\varTheta_{m,j}\right)\geq\left(2^{R_{req}}-1\right)\nonumber\\ & \left(P_j\varphi_{n,j}(|h_{m,j}|^2+\varTheta_{m,j}\right)+I_{m,j'}+\sigma^2), \forall j,\nonumber \\
 Z3:\ & P_j\varphi_{n,j}\leq P_j\varphi_{m,j},  \forall j,\nonumber\\
 Z4:\ & 0\leq P_j\leq P_{tot}, \forall j ,\nonumber\\
 Z5:\ & \varphi_{n,j}+\varphi_{m,j}\leq1, \forall j,\nonumber\\
 Z6:\ & 0\leq\beta_{j}\leq1, \forall i, \forall j,\nonumber
\end{alignat}
where $\boldsymbol{\varphi_j}=\{\varphi_{n,j},\varphi_{m,j}\}$. The constraints $Z1$ and $Z2$ guarantee the minimum rate of $U_n$ and $U_m$ in $C_j$, where $R_{req}$ is the minimum rate threshold. The transmit power of $C_j$ is controlled by $Z3$ and $Z5$, where $P_{tot}$ shows the total power budget of $C_j$. Moreover, constraint $Z4$ controls the power of $C_j$ while $Z6$ limits the RC of $D_j$. 
\begin{figure*}[!t]
\begin{align}
\phi_1 &= Q_{n,j} Q_{m,j}(-F_{m,j} Q_{n,j} (-1 + \varphi_{n,j})) (1+\varLambda_{m,j})+Q_{m,j}(F_{n,j} \varphi_{n,j} (1+\varLambda_{n,j})-Q_{n,j} \varPi_{1,j}),\tag{17}\label{17}\\
\phi_2&= Q_{n,j} Q_{m,j} (2 (H_{n,j} Q_{m,j} (-1+\varphi_{n,j}) -H_{m,j} Q_{n,j} \varphi_{n,j})\varPi_{1,j}+F_{m,j} (-1 + \varphi_{n,j}) (2 H_{n,j} (-1 + \varphi_{n,j})(1 + \varLambda_{m,j}) \nonumber\\& -F_{n,j} \varphi_{n,j} (2 + \varLambda_{n,j} + \varLambda_{m,j}) + Q_{n,j}\varPi_{1,j} ) +   F_{n,j}\varphi_{n,j} (2 H_{m,j} \varphi_{n,j} (1 + \varLambda_{n,j}) -Q_{m,j}\varPi_{1,j} ),\tag{18}\label{18}\\
\phi_3&=-(H_{n,j}^2 Q_{m,j}^2 (-1 + \varphi_{n,j})^2 - 4 H_{n,j}H_{m,j} Q_{n,j} Q_{m,j} (-1 +\varphi_{n,j}) \varphi_{n,j} +H_{m,j}^2Q_{n,j}^2 \varphi_{n,j}^2)\varPi_{1,j} \nonumber\\&+F_{n,j} \varphi_{n,j} (H_{m,j}^2 Q_{n,j} \varphi_{n,j}^2 (1 + \varLambda_{n,j}) + H_{n,j}Q_{m,j}^2 (-1 + \varphi_{n,j}) \varPi_{1,j} -  2 H_{m,j} Q_{n,j} Q_{m,j} \varphi_{n,j} \varPi_{1,j} ) \nonumber\\& - F_{m,j} (-1 + \varphi_{n,j})(H_{n,j}^2 Q_{m,j} (-1 + \varphi_{n,j})^2 (1 + \varLambda_{m,j}) - H_{n,j}Q_{m,j} (-1 +\varphi_{n,j}) (F_{n,j} \varphi_{n,j} (1 + \varLambda_{m,j})  \nonumber\\&- 2Q_{n,j}\varPi_{1,j} ) -Q_{n,j}\varphi_{n,j}(H_{m,j}Q_{n,j}\varPi_{1,j}  + F_{n,j} (-H_{m,j} \varphi_{n,j} (1 + \varLambda_{n,j}) +Q_{m,j} \varPi_{1,j} ))),\tag{19}\label{19}\\      
\phi_4&=(H_{m,j} \varphi_{n,j} (-2H_{n,j}^2Q_{m,j} (-1 + \varphi_{n,j})^2 + 2 H_{n,j} (H_{m,j}Q_{n,j}+ F_{n,j}Q_{m,j}) (-1 + \varphi_{n,j}) \varphi_{n,j}  \nonumber\\& -F_{n,j}H_{m,j}Q_{n,j} \varphi_{n,j}^2)+F_{m,j}(-1 + \varphi_{n,j}) (H_{n,j}^2 Q_{m,j} (-1 + \varphi_{n,j})^2 - H_{n,j} (2H_{m,j} Q_{n,j}  \nonumber\\&+F_{n,j}Q_{m,j})(-1 + \varphi_{n,j}) \varphi_{n,j} + F_{n,j} H_{m,j} Q_{n,j} \varphi_{n,j}^2))\varPi_{1,j} ,\tag{20}\label{20}\\
\phi_5&=-H_{n,j} H_{m,j} (-1 +\varphi_{n,j}) \varphi_{n,j} (H_{n,j} (-1 + \varphi_{n,j}) - F_{n,j} \varphi_{n,j}) (F_{m,j} - F_{m,j} \varphi_{n,j} + H_{m,j} \varphi_{n,j})\varPi_{1,j}.  \tag{21}   \label{21}
\end{align}\hrulefill
\end{figure*}

\section{Spectral Efficiency Optimization}
We can observe that problem (\ref{5}) is coupled on variable $\boldsymbol{\varphi_j}, P_j,\beta_{j}$. It makes this optimization problem non-convex and very hard to obtain the global optimal solution. Thus, (5) is divided into two subproblems and then we exploit the dual theory to get a sub-optimal yet efficient solution. For a given RC of $D_j$ in $C_j$, the subproblem of power allocation to maximize the system SE can be stated as
\begin{alignat}{2}
& \underset{{\boldsymbol{\varphi_j}, P_j}}{\text{max}} \sum\limits_{j=1}^J\{\log_2(1+\gamma_{n,j})+\log_2(1+\gamma_{m,j})\}\nonumber \\
&s.t. \quad (Z1)-(Z5),\label{6}
\end{alignat}
where $\gamma_{n,j}$ can also be written as
\begin{align}
\gamma_{n,j}=\frac{P_j\varphi_{n,j} F_{n,j}}{P_j\varphi_{m,j}{ H_{n,j}}+ Q_{n,j}}, 
\end{align}
with
${F_{n,j}}=|h_{n,j}|^2+G_{n,j}$ , ${H_{n,j}}=|h_{n,j}|^2\alpha$,
${Q_{n,j}}=I_{n,j'}+\sigma^2$. Accordingly, $\gamma_{m,j}$ can be stated as
\begin{align}
\gamma_{m,j}=\frac{P_j\varphi_{m,j}F_{m,j}}{P_j\varphi_{n,j}H_{m,j}+Q_{m,j}},
\end{align}
with
${F_{m,j}}=|h_{m,j}|^2+G_{m,j}$ ,
 ${H_{m,j}}=|h_{m,j}|^2+G_{m,j}$, and
${Q_{m,j}}=I_{m,j'}+\sigma^2$.
In the following, we prove that the $\log_2(1+\gamma_{n,j})+\log_2(1+\gamma_{m,j})$ in (\ref{6}) is concave with respect to $\boldsymbol{\varphi_j}$ and $P_j$. The dual problem is defined as $\underset{{\boldsymbol{\varLambda_{j}},\boldsymbol{\varPi_{j}}\ge0}}{\text{min}}\ D(\boldsymbol{\varLambda_{j}},\boldsymbol{\varPi_{j}})$, where $\boldsymbol{\varLambda_{j}}=\{\varLambda_{n,j},\varLambda_{m,j}\}$ and $\boldsymbol{\varPi_{j}}=\{\varPi_{1,j},\varPi_{2,j}\}$ represent the dual variables. Next we express the dual function as
\begin{align}
D(\boldsymbol{\varLambda_{j}},\boldsymbol{\varPi_{j}}) = \underset{{\boldsymbol{\varphi_{j}},P_j,\boldsymbol{\varLambda_{j}},\boldsymbol{\varPi_{j}}\ge 0}}{\text{max}} L(\boldsymbol{\varphi_{j}},P_j,\boldsymbol{\varLambda_{j}},\boldsymbol{\varPi_{j}}),\label{9}
\end{align}
where $L(\boldsymbol{\varphi_{j}},P_j,\boldsymbol{\varLambda_{j}},\boldsymbol{\varPi_{j}})$ is called the Lagrangian function,
\begin{alignat}{2}
& L(\boldsymbol{\varphi_{j}},P_j,\boldsymbol{\varLambda_{j}},\boldsymbol{\varPi_{j}})=\sum\limits_{j=1}^J\{\log_2(1+\gamma_{n,j})+\log_2(1+\gamma_{m,j})\}\nonumber\\
&+\varLambda_{n,j}(P_j\varphi_{n,j}F_{n,j}-(2^{R_{req}}-1){P_j\varphi_{m,j}{H_{n,j}}+Q_{n,j}})\nonumber\\
&+\varLambda_{m,j}(P_j\varphi_{m,j}F_{m,j}-(2^{R_{req}}-1)(P_j\varphi_{n,j}H_{m,j}\label{10}\\
&+Q_{m,j})+\varPi_{1,j}(P_{tot}-P_j)+\varPi_{2,j}(1-\varphi_{n,j}-\varphi_{m,j}).\nonumber
\end{alignat}
By adopting the KKT conditions and deriving (\ref{10}) with respect to $\varphi_{n,j}$, the power allocation coefficient can be calculated as
\begin{align}
&\varphi^*_{n,j}= \left(\frac {-A \pm \sqrt{A^2-4BS}}{2B}\right)^+, \\ 
&\varphi^*_{m,j}= 1- \varphi^*_{n,j}, 
\end{align}
where $(.)^+=\text{max}(0,.)$. The values of $A,B,S$ are stated in (\ref{13})--(\ref{15}) on the top of this page. Next we calculate the efficient $P_j$ in $C_j$ through deriving (\ref{10}) as
\begin{align}
\phi_1+\phi_2 P_{j}+\phi_3 P_{j}^2+\phi_4 P_{j}^3+\phi_5 P_{j}^4=0,\tag{16}\label{16}
\end{align}
where the values of $\phi_1,\phi_2,\phi_3,\phi_4,\phi_5$ are stated in (\ref{17})--(\ref{21}). We can see that (\ref{16}) is order four polynomial which can be efficiently solved through polynomial solver. With $\varphi^*_{n,j}, \varphi^*_{m,j}$ and $P^*_j$, the problem (\ref{9}) can be updated as

\begin{alignat}{2}
& \underset{{(\boldsymbol{\varphi^*_j}, P_{j}^*)}}{\text{max}} \sum\limits_{j=1}^J\bigg\{\log_2\bigg(1+\frac{P_j^*\varphi^*_{n,j}F_{n,j}}{P_j^*\varphi^*_{m,j}{H_{n,j}}+Q_{n,j}}\bigg)\nonumber\\
 &+\log_2\bigg(1+\frac{P_j^*\varphi^*_{m,j}F_{m,j}}{P_j^*\varphi^*_{n,j}H_{m,j}+Q_{m,j}}\bigg)\bigg\}\tag{22}\\
& s.t.\quad {\boldsymbol{\varLambda_j}, \boldsymbol{\varPi_j}\ge 0} \nonumber 
\end{alignat}
We calculate and iteratively update the Lagrangian variables as 
\begin{align}
\varLambda_{n,j}(t+1)&=\varLambda_{n,j}(t)+\xi(t)(P_j^*\varphi^*_{n,j}F_{n,j}\tag{23}\\
&-(2^{R_{req}}-1){P_j^*\varphi^*_{n,j}{H_{n,j}}+Q_{n,j}}), \forall j, \nonumber\\
\varLambda_{m,j}(t+1)&=\varLambda_{m,j}(t)+\xi(t)(P_j^*\varphi^*_{m,j}F_{m,j}\tag{24}\\
&-(2^{R_{req}}-1)(P_j^*\varphi^*_{n,j}H_{m,j}+Q_{m,j}),\forall j,\nonumber\\
\varPi_{1,j}(t+1)=&\varPi_{1,j}(t)+\xi(t)(1-(\varphi^*_{n,j}+\varphi^*_{m,j})),\forall j,\tag{25} \\
\varPi_{2,j}(t+1)=&\varPi_{2,j}(t)+\xi(t)(P_{tot}-P_{j}^*),\forall j,\tag{26}
\end{align}
where $\xi$ represents nonnegative step size. Equation (23)--(26) will iteratively updated until convergence criterion is satisfied.

Next we calculate the RC of $D_i$ in each cell. For the given $P_s, \varphi_{n,j}, \varphi_{m,j}$, the optimization problem (\ref{5}) can be reformulated as  
\begin{alignat}{2}
& \underset{{\beta_{j}}}{\text{max}} \sum\limits_{j=1}^J\{\log_2(1+\gamma_{n,j})+\log_2(1+\gamma_{m,j})\}\nonumber \\
&s.t. \quad (Z1), (Z2),(Z6),\tag{27}\label{27}
\end{alignat}
where $\gamma_{n,j}$ and $\gamma_{m,j}$ can be rewritten as $\gamma_{n,j}=(E_{n,j}+T_{n,j})/V_{n,j}$ with $E_{n,j}=P_j\varphi^*_{n,j}|h_{n,j}|^2, T_{n,j}=P_j\varphi^*_{n,j}G_{n,j}, V_{n,j}=P_j\varphi^*_{m,j}|h_{n,j}|^2\alpha+I_{n,j'}+\sigma^2$, and $\gamma_{m,j}=(E_{m,j}+T_{m,j})/(V_{m,j}+\eta_{m,j})$ with $E_{m,j}=P_j\varphi^*_{m,j}|h_{m,j}|^2, T_{m,j}=P_j\varphi^*_{m,j}G_{m,j}, V_{m,j}=P_j\varphi^*_{n,j}|h_{m,j}|^2+I_{m,j'}+\sigma^2, \eta_{m,j}=P_j\varphi^*_{n,j}G_{m,j}$. Again, $\log_2(1+\gamma_{n,j})+\log_2(1+\gamma_{m,j})$ in (\ref{27}) is concave with respect to $\beta_{i,j}$. To obtain the efficient solution, we employ the KKT conditions, where the closed-form expression of $\beta_{j}$ is derived as
\begin{align}
\beta^*_{j}= \left(\frac {(2^{R_{req}}-1)-E_{n,j}}{T_{n,j}}\right), \tag{28}\label{28}
\end{align}
Now using the alternate optimization, we first calculate the value of transmit power by fixing RC at BD. Then, for a given value of the power, we calculate RC at BD. The framework of optimizing the two sub-problems is also presented in Algorithm 1.
\begin{algorithm}[t]
{\bf Initialize:} we first initialize all the system parameters
 
 {\bf Step 1:} calculate $\varphi_{n,j}$, $\varphi_{m,j}$, and $P_j$ for the given $\beta_j$
         
    \While{not converge}{\For{$j=1:J$, $n=1:I$, $m=1:J$}{Compute $\varphi_{n,j}$ and $\varphi_{m,j}$ using (11) and (12) \\
        Compute $P_j$ by (16)\\
        Iteratively update (23)--(26) }}
{\bf Step 2:} Now with $\varLambda^*_{i,k},\varLambda^*_{i,k},P^*_k$ in (27), we compute $\beta_j$
    
    \For{$j=1:J$}{Calculate $\beta_j$ using Equation (28) 
    }

    Return $\varphi^*_{n,j}$, $\varphi^*_{m,j}$, $P^*_j$,  $\beta^*_{j}$
    \caption{Framework of optimizing (6) and (27)}
   \end{algorithm}  

\section{Results and Discussion}
In this section, we provide the numerical results based on Monte Carlo simulations. We compare the proposed NOMA-BC scheme and the benchmark NOMA scheme with no backscattering (NOMA-NB). Unless mentioned otherwise, the simulation parameters are: the wireless channels undergo Rayleigh fading, the power of BS is $P_{tot}=30$ dBm, the number of cells is $J=5$, the RC of a $D_j$ is $0\leq\beta_{j}\leq1$, the path-loss exponent is 3, the parameter of imperfect signal decoding varies from 0.1 to 1, and $\sigma^2=0.1$. Moreover we consider two NOMA users and one BD in each cell for simplicity.   

\begin{figure}[!t]
\centering
\includegraphics [width=0.49\textwidth]{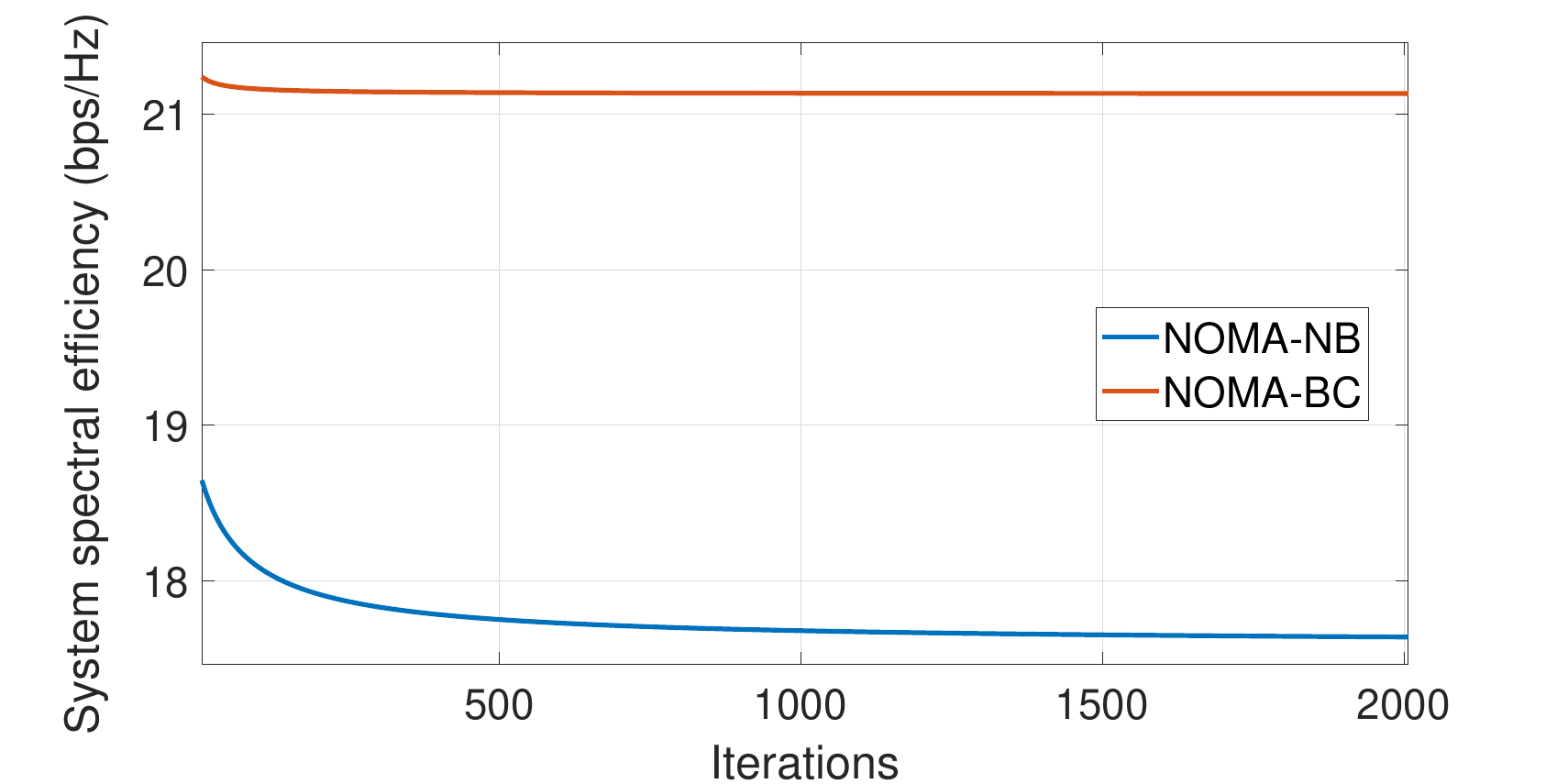}
\caption{Number of iteration versus total SE of NOMA-BC and NOMA-NB schemes, where $P_{tot}=30$ dBm.}
\label{convergence}
\end{figure}
Fig. \ref{convergence} shows the convergence of the proposed NOMA-BC scheme provided in Section III and the benchmark NOMA-NB scheme with efficient power allocation. We plot the system SE against the number of iterations. It can be seen that both schemes converge within limited iterations. It is worth-mentioning here to discuss the complexity which can be computed based on the number of iterations. Specifically, the complexity of the proposed scheme in each iteration is calculated as $\mathcal O\{2J\}$. If the total number of iterations is $T$, then, the total complexity can be expressed as $\mathcal O\{T2J\}$.

\begin{figure}[!t]
\centering
\includegraphics [width=0.49\textwidth]{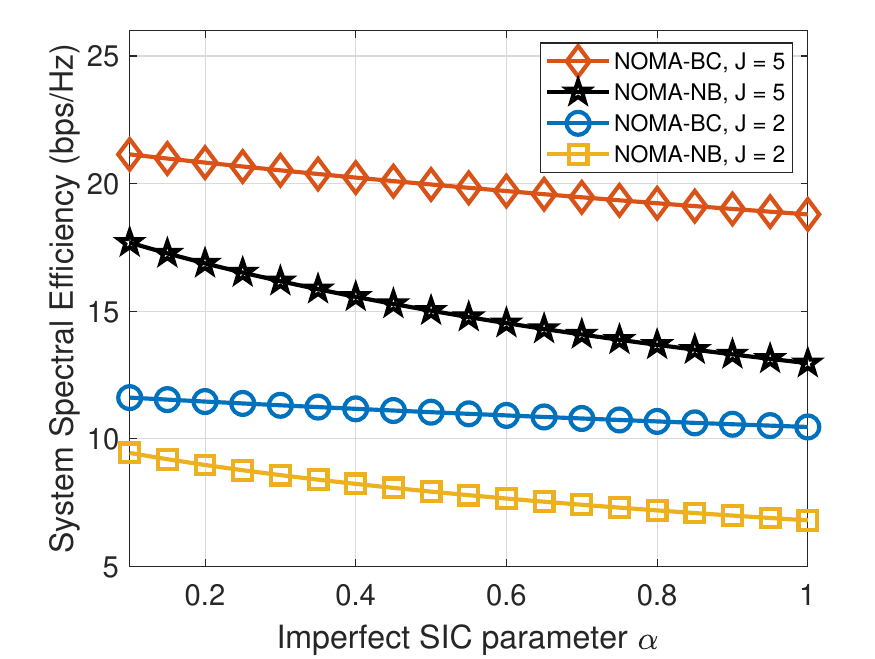}
\caption{The effect of increasing $\alpha$ on the SE with different $J$ values with $P_{tot}=30$ dBm and $R_{req}=0.5$ bps/Hz.}
\label{iSICvsSEvsJ}
\end{figure}

Next, we plot and discuss the impact of signal decoding errors on the performance of the system. Fig. \ref{iSICvsSEvsJ} depicts the SE of the system versus varying values of $\alpha$ for the different numbers of cells in the system. In both cases, we can see that NOMA-BC scheme performs better than the conventional NOMA-NB scheme. Moreover, both schemes perform well for the small values of $\alpha$. However, an increase in $\alpha$ decreases the total SE of both schemes. The increase in $\alpha$ reduces users' signal decoding capability, resulting in interference to the system.

\begin{figure}[!t]
\centering
\includegraphics [width=0.49\textwidth]{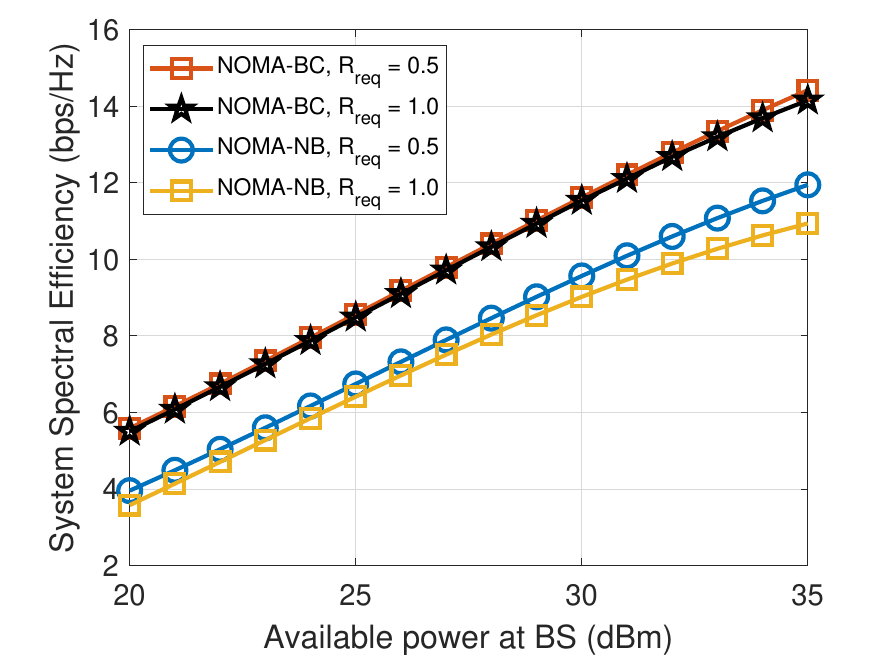}
\caption{The impact of $P_{tot}$ on the SE of multi-cell network with different $R_{req}$, $J=2$, $\alpha=0.5$.}
\label{PvsSEvsRmin}
\end{figure}

Fig. \ref{PvsSEvsRmin} shows the impact of the total power on the system performance by plotting the system SE against $P_{tot}$, where $J=2$ and the minimum required rate is set as $R_{req}=0.5$ and $R_{req}=1.0$ bps/Hz. It is evident that the increase in $P_{tot}$ enhances the total SE because more transmit power becomes available for transmission. Moreover, this plot also show the effect of $R_{req}$ on the overall system performance. The increase in $R_{req}$ results in reduced SE because more resources are required to meet the high $R_{req}$ of users. It is evident that the proposed NOMA-BC scheme provides better performance than the NOMA-NB scheme. This is because, in the proposed scheme, users also benefit from the additional gain of BC along with the optimization of all other resources. 

\section{Conclusions}
The combination of BC and NOMA is expected to support massive connection of low-powered wireless devices in B5G. This paper has provided a new optimization approach to enhance the SE of NOMA-BC network under imperfect signal decoding. More specifically, the power allocation of BS and RC of BD are simultaneously optimized in each cell under various practical constraints. The original problem of SE maximization has been divided into two subproblems first. Then, efficient solutions have been carried through the decomposition method and KKT conditions. Simulation results have confirmed the superiority of the proposed NOMA-BC compared to the benchmark NOMA framework. In future, we plan to extend this work to a multi-carrier communications. In such a case, the system performance can be further improved by optimizing user association and subcarrier assignment.

\bibliographystyle{IEEEtran}
\bibliography{Wali_EE}

\end{document}